\documentclass[nonacm,sigconf,natbib]{acmart}

\usepackage{adjustbox}
\usepackage{array,booktabs}
\usepackage{amsmath, amsfonts, amsthm}
\usepackage{bm}
\usepackage{graphicx}
\graphicspath{{./}{Charts/}}
\usepackage{subcaption}
\usepackage{mathtools}
\usepackage{balance}

\usepackage{array}
\newcolumntype{H}{>{\setbox0=\hbox\bgroup}c<{\egroup}@{}}

\newcommand{\commentout}[1]{}
\newcommand{\eqnref}[1]{Eq.~(\ref{#1})}
\newcommand{\eqnsref}[2]{Eqs.~(\ref{#1}) and (\ref{#2})}

\newcommand{\figref}[1]{Figure~\ref{#1}}

\newcommand{\tblref}[1]{Table~\ref{#1}}

\DeclareSymbolFont{libertineg}{\encodingdefault}{\familydefault}{m}{it}
\SetSymbolFont{libertineg}{bold}{\encodingdefault}{\familydefault}{b}{it}
\DeclareMathSymbol{g}{\mathalpha}{libertineg}{`g}

\makeatother

\begin{document}

\title{Learning to Rank when Grades Matter}

\author{%
Le Yan, Zhen Qin, Gil Shamir, Dong Lin, Xuanhui Wang, Mike Bendersky
}
\affiliation{ 
 \institution{Google}
  \streetaddress{1600 Amphitheater Pkwy}
  \city{Mountain View} 
  \state{CA 94043}
}
\email{{lyyanle, zhenqin, gshamir, dongl, xuanhui, bemike}@google.com}

\begin{abstract}
Graded labels are ubiquitous in real-world learning-to-rank applications, especially in human rated relevance data. Traditional learning-to-rank techniques aim to optimize the ranked order of documents. They typically, however, ignore predicting actual grades. This prevents them from being adopted in applications where grades matter, such as filtering out ``poor'' documents. Achieving both good ranking performance and good grade prediction performance is still an under-explored problem. Existing research either focuses only on ranking performance by not calibrating model outputs, or treats grades as numerical values, assuming labels are on a linear scale and failing to leverage the ordinal grade information.
In this paper, we conduct a rigorous study of learning to rank with grades, where both ranking performance and grade prediction performance are important. We provide a formal discussion on how to perform ranking with non-scalar predictions for grades, and propose a multiobjective formulation to jointly optimize both ranking and grade predictions. In experiments, we verify on several public datasets that our methods are able to push the Pareto frontier of the tradeoff between ranking and grade prediction performance, showing the benefit of leveraging ordinal grade information.

\end{abstract}

\maketitle

\fancyhead{}

\section{Introduction} \label{sec:intro}
Learning to rank (LTR) with graded labels is ubiquitous in real-world applications. For example, in traditional LTR datasets such as Web30K, human raters rate each query-document pair from ``irrelevant'' (graded as 0) to ``perfectly relevant'' (graded as 4). Grades are \emph{ordinal\/}, i.e., represented by discrete numbers with a natural order, but not necessarily preserving numerical relations. For example, grade 4 is not necessarily twice as relevant as grade 2. 
Traditional LTR work focuses on ranking performance or treats grades as numerical values~\cite{yan2022scale}, ignoring potential non-linearity of the grading scale. Predicting actual grades is traditionally treated as a classification problem, which has not been given much attention in the LTR literature~\cite{li2007mcrank}, and that usually ignores the order of the grades. Unlike classical LTR work, we consider the problem in which both ranking performance and grade prediction performance, measured by ranking metrics and classification accuracy, respectively, are both important. We argue that achieving good performance on both fronts delivers a better user facing experience via optimal ranking \emph{and} capabilities such as filtering out ``poor'' documents with certain grades. For example, user could choose to show just perfectly relevant results or any relevant results when grade predictions are available. 

In the sequel, we present a rigorous study of LTR with graded labels. 
We formally demonstrate ranking with non-scalar predictions for grades. Based on ordinal prediction aggregation, we propose a multiobjective formulation that directly trades-off ranking and grade prediction. We conduct an extensive experimental study on 3 public LTR datasets, comparing with state-of-art ranking methods, and ranking-agnostic classification methods. Experimental results show interesting trade-off behaviors of different methods.  Our proposed methods are able to push the Pareto frontier of ranking and grade prediction performances.  



\section{Related Works} \label{sec:related}
LTR has been widely studied with focus on designing losses and optimization methods to improve ranking performance. Several notable losses include Pairwise Logistic~\cite{burges2005learning} (also called RankNet) and ListNet~\cite{xia2008listwise}. Subsequent work included multiple perspectives to optimize ranking metrics.  These include LambdaRank~\cite{burges2010ranknet}, SoftNDCG~\cite{taylor2008softrank}, SmoothNDCG~\cite{chapelle2010gradient}, ApproxNDCG~\cite{qin2010general, bruch2019revisiting} and GumbelApproxNDCG~\cite{bruch2020stochastic}, among many others.
A recent work (LambdaLoss~\cite{jagerman2022optimizing}) used ideas from LambdaRank to develop a theoretically sound framework for neural optimization of ranking metrics.

Ranking methods studied in the LTR literature focus on improving ordering, but not on prediction accuracy of the actual labels (or grades). 
Previous work~\cite{li2007mcrank} studied if accurate label predictions could lead to good ranking, but not directly optimizing both objectives. Multi-objective setting has been well studied in Gradient Boosting Decision Trees~\cite{chapelle2010multi, svore2011learning, carmel2020multi}, but little attention has been paid to the two objectives we are considering.
Calibrated LTR, where model predictions are anchored to concrete meanings, has also drawn some attention due to its practical value~\cite{CTR:LTR:ADKDD13, Calibrate:ADKDD17, yan2022scale}. However, existing work treats grades as real values, assuming that grade values are on a linear scale. This is inaccurate for many applications where the grades are ordinal and discrete, but not linear. To the best of our knowledge, our work is the first to formally study and demonstrate benefit on various tasks of learning to rank with graded labels when prediction of the labels matter. 

\section{Problem Formulation} \label{sec:problem}
We consider a ranking dataset with graded documents, 
\[
\mathcal{D}=\{\{q, \{{\bf x}_i, y_i\}|i\in\mathcal{D}_q\}|q\in\mathcal{Q}\}.
\]
Dataset $\mathcal{D}$ consists of queries $q \in \mathcal{Q}$, each associated with a set of candidate documents $\mathcal{D}_q$.  Document $i$ is featured by ${\bf x}_i$ and graded label $y_i$.  Without loss of generality, we assume $y_i\in\{0, 1, ..., L-1\}$ for $L$ possible ordinal classes. The ordinal relevance relation aligns with the integer order.
The graded labels in the setting play two aligned roles: (1) they define the ordinal categories that a document appears in a query; and (2) presenting the list of documents in descending order of the grades optimizes ranking performance.

Conventionally, optimization focuses on one of two objectives: (1) to predict the correct category of each query document pair; or (2) to exploit the correct ranking regardless of the category predictions. 
Ideally, as the perfect ranking can be achieved by sorting the grades, i.e., perfect category predictions indicate perfect ranking, optimizing (1) is sufficient to reach (2). In practice, however, directly optimizing (2) usually leads to much better ranking performance. 
In this work, we consider a formulation to jointly optimize the two objectives. 

\subsection{Ordinal grade prediction}
We aim 
to predict correct graded labels for query-document pairs.

\paragraph{Mean squared error.}
A naive and straight-forward way is to cast ordinal classes as real values and then apply linear regression. We consider a parametric model that predicts a real value for each query-document pair minimizing mean squared error between the model prediction $f_\theta({\bf x}_i)$ and the graded label $y_i$,
\begin{equation}
    \label{eq:mse}
    \mathcal{L}^{\rm MSE}(\mathcal{D}) = \sum_i\left(f_\theta({\bf x}_i) - y_i \right)^2.
\end{equation}
The model converges to the expected $y_i$, and we can pick the grade that minimizes the distance to the model's prediction,
\begin{equation}
    \label{eq:mselabel}
    \hat{y}_i = {\rm argmin}_{l=0, ..., L-1} |l - f_\theta({\bf x}_i)|.
\end{equation}
An implicit assumption is that the grade scale is well calibrated.  Thus, differences in relevance are equal if differences in labels are equal.  However, this may not be the case for every graded dataset.

\paragraph{Multi-class cross entropy.}
Making predictions of graded categories can be seen as a multi-class classification problem, and the presumption above is no longer needed. 
The model predictions, ${\bf f}_\theta({\bf x}_i)$, with $L$ logits for $L$ grades, can be transformed to normalized probabilities with a softmax function,
\begin{equation}
    p(y_i = l| {\bf x}_i) = \frac{\exp({ f}^l_\theta({\bf x}_i))}{\sum_j\exp({f}^j_\theta({\bf x}_i))}.
    \label{eq:prob}
\end{equation}
The superscript $l$ labels the $l$-th component of the predictions. The model is trained to minimize cross-entropy loss,
\begin{equation}
    \label{eq:mcce}
    \mathcal{L}^{\rm CE}(\mathcal{D}) = -\sum_i\sum_{l=0}^{L-1}\mathbb{I}(y_i=l) \ln(p(y_i = l| {\bf x}_i)),
\end{equation}
where $\mathbb{I}(y_i=l)$ is the indicator function of item $i$ taking label $l$.
Given the model predicted probabilities of each ordinal category, we naturally use the label maximizing the corresponding probability as the predicted ordinal grade,
\begin{equation}
    \label{eq:problabel}
    \hat{y}_i = {\rm argmax}_{l=0, ..., L-1}p(y_i = l| {\bf x}_i).
\end{equation}



The multi-class cross entropy approach ignores the ordinal relation of grades, which could possibly be leveraged in training.  For example, if a document is not likely in grade $l$ or higher, then it is less likely in grade $l+1$ or higher. Ordinal regression methods have been applied to leverage this relation.


\paragraph{Univariate ordinal regression.}
Univariate ordinal regression leverages ordinal relations by mapping ordinal grades into consecutive regions on the real axis. $L-1$ variables $\phi_1$, $\phi_2$, $...$, $\phi_{L-1}$, constrained to $\phi_l\leq \phi_m$ {\it iff} $l<m$, are trained as class boundaries for the full dataset (or slices of it). Together with $\phi_0=-\infty$ and $\phi_L=\infty$, the $L+1$ boundaries partition the real axis into $L$ consecutive regions. A model learns a per-item shift $f_\theta(\mathbf{x}_i)$ for the grid of boundaries.  Fitting the shifted boundaries to an infinite support \emph{probability density function (PDF)} renders the integral over each region as the class probability (where integrating from $-\infty$ to a shifted boundary gives the \emph{cumulative density function (CDF)} of an item up to some label class).  Fitting a \emph{logistic PDF} gives probability,
\begin{equation}
    \label{eq:univariate}
    p(y_i\geq l|{\bf x}_i) = \frac{1}{1+\exp(-[f_\theta({\bf x}_i) - \phi_l])}
\end{equation}
for item $i$ belonging to class $l$ or greater. Thus, 
\begin{multline}
    \label{eq:uniprob}
    p(y_i = l|{\bf x}_i) = p(y_i\geq l|{\bf x}_i) - p(y_i\geq l+1|{\bf x}_i) \\
    = \frac{1}{1+\exp(-[f_\theta({\bf x}_i) - \phi_l])} - \frac{1}{1+\exp(-[f_\theta({\bf x}_i) - \phi_{l+1}])}
\end{multline}
is the probability of $i$ taking label $l$.
With the probability in \eqnref{eq:uniprob}, the model $f_\theta$ and boundaries $\{\phi_l\}$ are trained to minimize the cross entropy loss in \eqnref{eq:mcce}.

\paragraph{Multivariate ordinal regression.}
Multivariate ordinal regression, see also in Ref.~\cite{pobrotyn2020context}, leverages the ordinal relations by dividing the $L$-level ordinals into $L-1$ successive binary classifications, which learn $L-1$ values $f^l_\theta({\bf x}_i)$, each with logistic regression, giving  
\begin{equation}
    \label{eq:multivariate}
    p(y_i\geq l | {\bf x}_i) = \frac{1}{1+\exp(-f^l_\theta({\bf x}_i))},\quad{\rm for}\ l = 1, 2, ..., L-1,
\end{equation}
with $p(y_i\geq 0 | {\bf x}_i) = 1$ and $p(y_i\geq L | {\bf x}_i) = 0$.
Then, 
\begin{equation}
    \label{eq:multiprob}
    p(y_i = l | {\bf x}_i) = \frac{1}{1+\exp(-f^l_\theta({\bf x}_i))} - \frac{1}{1+\exp(-f^{l+1}_\theta({\bf x}_i))}.
\end{equation}

The multivariate ordinal regression trains the model to minimize 
the sum of the $L-1$ consecutive logistic losses,
\begin{multline}
    \label{eq:vanillaordinal}
    \mathcal{L}^{\rm Ord}(\mathcal{D}) = -\sum_i\sum_{l=1}^{L-1}\left[\mathbb{I}(y_i\geq l) \ln(p(y_i\geq l | {\bf x}_i)) \right.\\
    \left.+ \mathbb{I}(y_i<l) \ln(1 - p(y_i\geq l | {\bf x}_i))\right].
\end{multline}

Both univariate and multivariate ordinal methods could predict the grade using max probability, as in \eqnref{eq:problabel}.

\begin{table*}
\centering
\caption{The statistics of the three largest public benchmark datasets for LTR models.}%
\vspace{-0.3cm}
\label{tbl:datasets}
\resizebox{0.6\linewidth}{!}{
\begin{tabular}{c|c|ccc|c|ccccc}
\hline & \#features & \multicolumn{3}{c|}{\#queries} & avg.  & \multicolumn{5}{c}{grade ratio (\%)} \\
 &  & training & validation & test &  \#docs  & 0 & 1 & 2 & 3 & 4 \\
\hline
Web30K & 136 & 18,919 & 6,306 & 6,306 & 119 & 51.4 & 32.5 & 13.4 & 1.9 & 0.8\\

Yahoo & 700 & 19,944 & 2,994 & 6,983 & 24   & 26.1 & 35.9 & 28.5 & 7.6 & 1.9\\

Istella & 220 & 20,901 & 2,318 & 9,799 & 316 & 96.3 & 0.8 & 1.3 & 0.9 & 0.7\\
\hline
\end{tabular}
}
\vspace{-.1in}
\end{table*}



\subsection{Ranking prediction}
\label{sec:ranking}
LTR methods usually care only about the ranking of documents in the same list, and can be insensitive to the absolute values of predictions.
In the most popular state-of-the-art ranking methods, as introduced below, the model predicts a ranking score,  $s_i = f_\theta({\bf x}_i)\in\mathbb{R}$, for each query-document pair, and the documents in the same query list are then ranked by sorting their scores.

\paragraph{Lambda loss.}
As ranking performance is usually measured by ranking metrics, some methods directly optimize these metrics or corresponding surrogates. The Lambda loss~\cite{burges2010ranknet,jagerman2022optimizing} is an example, 
where we reweight the gradient of each pair in a pairwise logistic loss ~\cite{burges2005learning} by the difference between the ranking metric to its value when flipping the pair.
To optimize the \emph{Normalized Discounted Cumulative Gain (NDCG)} metric \cite{jarvelin2002cumulated}, we apply 
\begin{equation}
    \label{eq:lambda}
    \mathcal{L}^{\rm Lambda}(\mathcal{D})
    = -\sum_{q\in\mathcal{Q}}\sum_{i, j\in \mathcal{D}_q: y_i > y_j}\Delta_{i,j}\ln\frac{1}{1+\exp(-(s_i - s_j))},
\end{equation} 
where $\Delta_{ij}$ is the LambdaWeight as defined in Eq. (11) of \cite{jagerman2022optimizing}.

\section{Methods} \label{sec:methods}
The main challenge to balance the two roles of the graded labels is to align grade prediction methods and ranking methods.

\paragraph{Ranking score of grade prediction methods.}
Compared with \textit{mean squared error} and \textit{univariate ordinal} methods, where we can directly leverage the scalar predictions as the ranking scores, $s_i=f_\theta({\bf x}_i)$, it is less straightforward to determine ranking scores for the multivariate \textit{multi-class cross entropy} and \textit{ordinal} methods.
The multivariate output corresponds to well-defined probabilities as in \eqnsref{eq:prob}{eq:multivariate}, but contains only part of the information for ranking. A single output scalar is insufficient for ranking. 
We thus propose to use the expected grade predictions in these methods as the ranking scores, which align with sorting by grades.
Following \eqnref{eq:prob}, for \textit{multi-class cross entropy} method, we have
\begin{equation}
    \label{eq:mcceexpect}
    s_i=\mathbb{E}[y_i] = \sum_{l=0}^{L-1}lp(y_i=l|{\bf x}_i) = \sum_{l=0}^{L-1}l\frac{\exp(f_\theta^l({\bf x}_i))}{\sum_j\exp(f_\theta^j({\bf x}_i))}.
\end{equation}
Following \eqnref{eq:multivariate}, assuming equally spaced consecutive label values, for the \textit{multivariate ordinal} method, we have
\begin{equation}
    \label{eq:ordexpect}
    s_i=\mathbb{E}[y_i] = \sum_{l=1}^{L-1}[l - (l-1)]p(y_i\geq l|{\bf x}_i) = \sum_{l=1}^{L-1}\frac{1}{1+\exp(-f_\theta^l({\bf x}_i))}.
\end{equation}

\paragraph{Multiobjective methods}
Given the ranking score from the ordinal predictions in \eqnsref{eq:mcceexpect}{eq:ordexpect}, we can also extend the multiobjective setting to \textit{multi-class cross entropy} and \textit{multivariate ordinal} methods, with a total loss,
\begin{equation}
    \label{eq:multiobj}
    \mathcal{L}^{\rm MultiObj}(\mathcal{D}) = (1-\alpha) \mathcal{L}^{\rm Ord}(\mathcal{D}; f_\theta) + \alpha \mathcal{L}^{\rm Rank}(\mathcal{D}; s),
\end{equation}
where the ranking score function $s$ is defined by the grade prediction function $f_\theta$, and $\alpha$ gives the relative weight on the ranking method.

\section{Experiments} \label{sec:exp}


\subsection{Experimental Setup}

We study the problem with three large public learning-to-rank datasets, Web30K~\cite{qin2013introducing}, Yahoo~\cite{yahoo}, and Istella~\cite{dato2016fast}. The statistics of the datasets used are summarized in \tblref{tbl:datasets}.




\paragraph{Comparing Methods}
The focus of this paper is on the loss function, thus all compared methods on each dataset share the same model architecture, containing three layers with 1024, 512, 256 hidden units, implemented with a public learning to rank library: TensorFlow Ranking~\footnote{https://github.com/tensorflow/ranking}. In addition, we apply the log1p input transformations, batch normalization, and dropout~\cite{dasalc}. Hyperparameters including learning rate, batch normalization momentum, dropout rate, and rank loss weight $\alpha$ are tuned for each method when applicable to the validation set. 

As summarized in \tblref{tbl:methods}, we study the naive methods (MSL, MCCE, UniOrd, and Ordinal) that train models to directly predict relevance grades, compared with the SOTA ranking methods (Lambda, Softmax, USoft, Gumbel), as well as the multiobjective methods allowing us to optimize both grade prediction accuracy and ranking simultaneously.
\begin{table}[t]
\caption{Compared methods.}
\label{tbl:methods}
\resizebox{\columnwidth}{!}{
\begin{tabular}{l|p{0.36\textwidth}}
\toprule
    Method & Description \\
    \midrule
    MSL  & Mean squared error loss method in \eqnref{eq:mse}. \\
    MCCE & Multi-class classification in \eqnref{eq:mcce}. \\
    UniOrd & Univariate Ordinal regression in \eqnref{eq:uniprob}. \\
    Ordinal & Vanilla multivariate Ordinal regression in \eqnref{eq:multivariate}. \\
    Lambda~\cite{jagerman2022optimizing} & LambdaLoss@1 method optimizing NDCG metric in \eqnref{eq:lambda}. \\
    MSL (Lambda)~\cite{yan2022scale} & Multiobjective method combining MSL and Lambda in \eqnref{eq:multiobj}. \\
    MCCE (Lambda) & Multiobjective method combining MCCE and Lambda in \eqnsref{eq:mcceexpect}{eq:multiobj}. \\
    UniOrd (Lambda) & Multiobjective method combining UniOrd and Lambda in \eqnref{eq:multiobj}. \\
    Ordinal (Lambda) & Multiobjective method combining Ordinal and Lambda in \eqnsref{eq:ordexpect}{eq:multiobj}. \\
\bottomrule
\end{tabular}
}
\vspace{-.2in}
\end{table}

\begin{table*}[t]
\caption{Comparisons on classification and ranking for three LTR datasets. 
Bold numbers are the best in each column. Up arrow ``$\uparrow$'' and down arrow ``$\downarrow$'' indicate statistical significance with p-value=0.01 of better and worse ACC/NDCG performance than the multiobjective baseline ``MSL (Lambda)'', respectively. 
The results of multiobjective methods in the table correspond to the ones of optimal balance of ACC and NDCG@10, as the bold markers in ~\figref{fig:tradeoff}.}
\vspace{-.05in}
\label{tbl:result}
\resizebox{\linewidth}{!}{
\begin{tabular}{cccccHccccccccH}
\hline
 & \multicolumn{5}{c}{Web30K} & \multicolumn{4}{c}{Yahoo} & \multicolumn{5}{c}{Istella}\\
\hline Method & CE & MSE & ACC & NDCG@10 & OrdAUC & CE & MSE & ACC & NDCG@10 & CE & MSE & ACC & NDCG@10 & OrdAUC \\
\hline
\hline 
{MSL}      & 12.348          & 0.5414 & 0.5531$^\uparrow$      & 0.5002$^\downarrow$     & 0.6634  & 13.570 & 0.5781 & 0.5089 & 0.7720 & 1.8310           & {0.1166} & 0.9337$^\downarrow$  & 0.7120$^\downarrow$   & 0.9655 \\
\hline
{MCCE}     & 0.9035           & 0.5384  & 0.6018$^\uparrow$      & 0.5028$^\downarrow$  & 0.6341  & \textbf{1.0531} & \textbf{0.5736} & \textbf{0.5260}$^\uparrow$ &  0.7722        & \textbf{0.1236}           & \textbf{0.1085} & 0.9611$^\uparrow$  & 0.7111$^\downarrow$     & 0.9661 \\
{UniOrd}   & 0.9202             & 1.5899 & 0.5953$^\uparrow$    & 0.4953$^\downarrow$ & 0.6645  & 1.0916 & 1.6029 & 0.5155$^\uparrow$ & 0.7692$^\downarrow$ & 0.1276 & 112.39 & 0.9612$^\uparrow$ & 0.7151$^\downarrow$ & 0.9697$^\uparrow$ \\
{Ordinal}  & 0.9066  & 0.5405 & {0.6013}$^\uparrow$  & 0.5053   & 0.6699 & {1.0628} & {0.5763} & {0.5235}$^\uparrow$ & 0.7698$^\downarrow$ & {0.1252} & {0.1093} & \textbf{0.9616}$^\uparrow$ & 0.7123$^\downarrow$ & 0.9680$^\uparrow$ \\
\hline
{Lambda}~\cite{jagerman2022optimizing}   & 0.9444 & 1.8466 & 0.5709$^\uparrow$ & 0.5057   & 0.6626$^\uparrow$  & 1.4078 & 3.8040 & 0.2993$^\downarrow$ & 0.7716 & 0.1577 & 544.92 & 0.9578$^\uparrow$ & {0.7310}$^\uparrow$ & 0.9684$^\uparrow$ \\
\hline
MSL (Lambda)~\cite{yan2022scale}   & 12.543 & 0.5566 & 0.5460      & 0.5054 & 0.6703  & 13.591 & 0.5781 & 0.5081 & {0.7726} & 1.6886            & 0.1215 & 0.9389           & 0.7251  & \textbf{0.9706}$^\uparrow$ \\
\hline
MCCE (Lambda)   & \textbf{0.9027} & \textbf{0.5377} & \textbf{0.6030}$^\uparrow$  & \textbf{0.5107}$^\uparrow$ & 0.6703  & 1.0604 & \textbf{0.5736} & 0.5232$^\uparrow$ & 0.7734 & 0.1328 & 0.1206 & 0.9605$^\uparrow$ & 0.7288$^\uparrow$ & 0.9706$^\uparrow$ \\
UniOrd (Lambda)   & 0.9280 & 1.5973 & 0.5877$^\uparrow$  & 0.5073 & 0.6703  & 1.1109 & 1.5923 & 0.5040 & 0.7721 & 0.1422 & 319.43 & 0.9581$^\uparrow$ & \textbf{0.7320}$^\uparrow$ & 0.9706$^\uparrow$ \\
Ordinal (Lambda)   & {0.9056} & {0.5394} & 0.6006$^\uparrow$  & {0.5100}$^\uparrow$ & 0.6703  & 1.0650 & 0.5758 & 0.5225$^\uparrow$ & \textbf{0.7743}$^\uparrow$ & 0.1365 & 0.1242 & 0.9593$^\uparrow$ & 0.7298$^\uparrow$ & \textbf{0.9706}$^\uparrow$ \\
\hline
\end{tabular}
}
\end{table*}

\begin{figure*}[tbp]
  \centering
  (a) Web30K\hspace{0.25\linewidth}  (b) Yahoo\hspace{0.25\linewidth} (c) Istella\\
  \includegraphics[width=.32\linewidth]{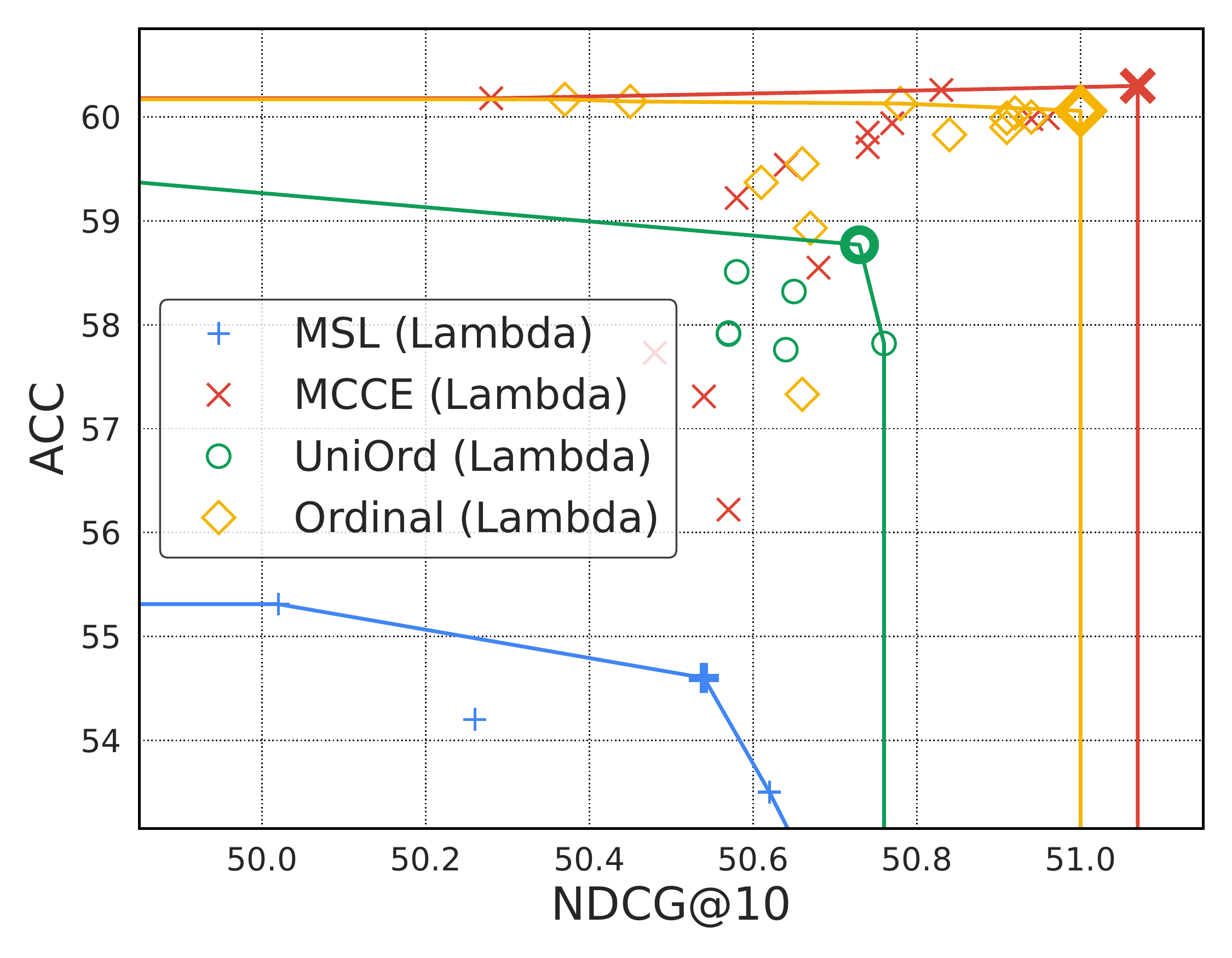}
  \includegraphics[width=.32\linewidth]{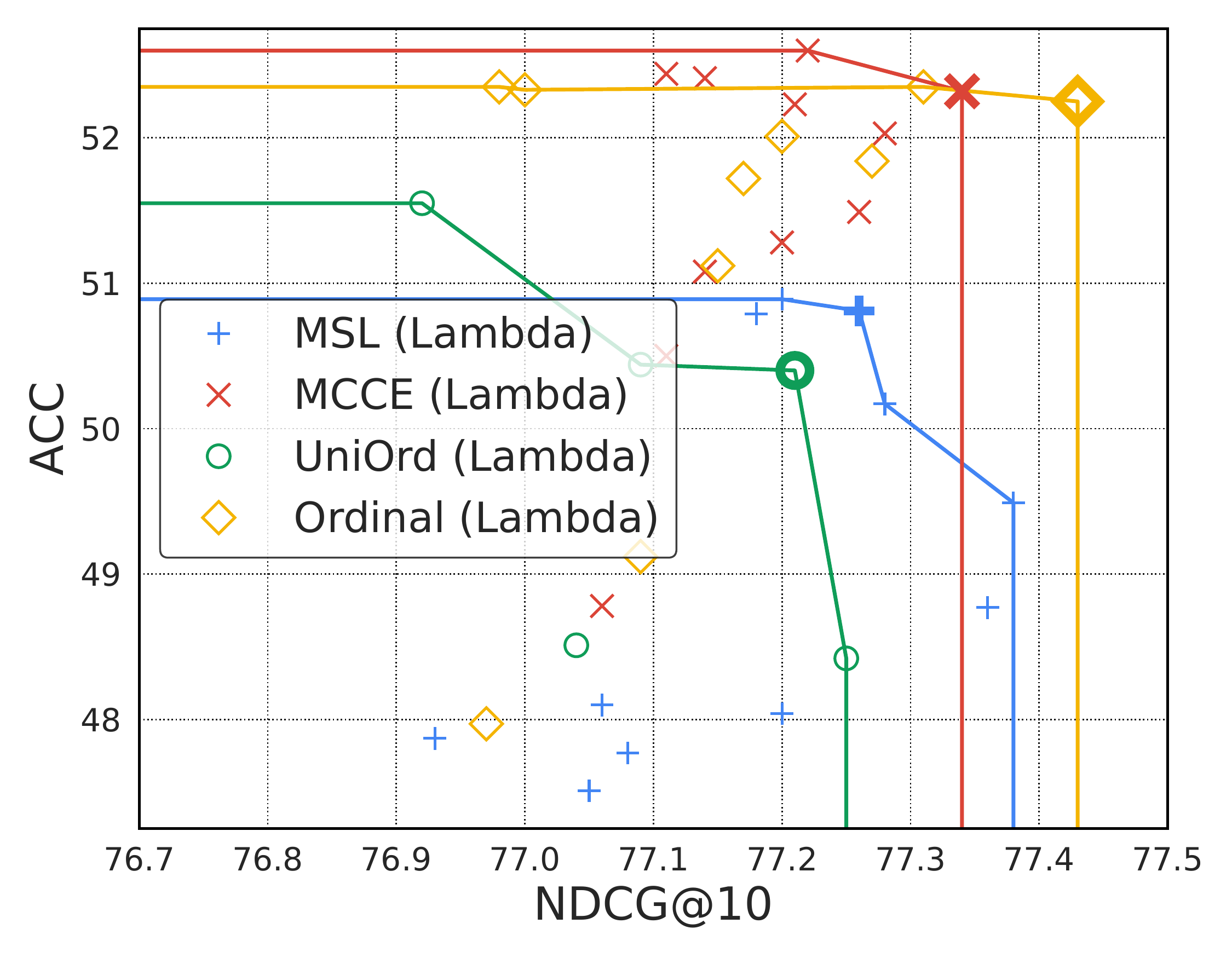}
  \includegraphics[width=.32\linewidth]{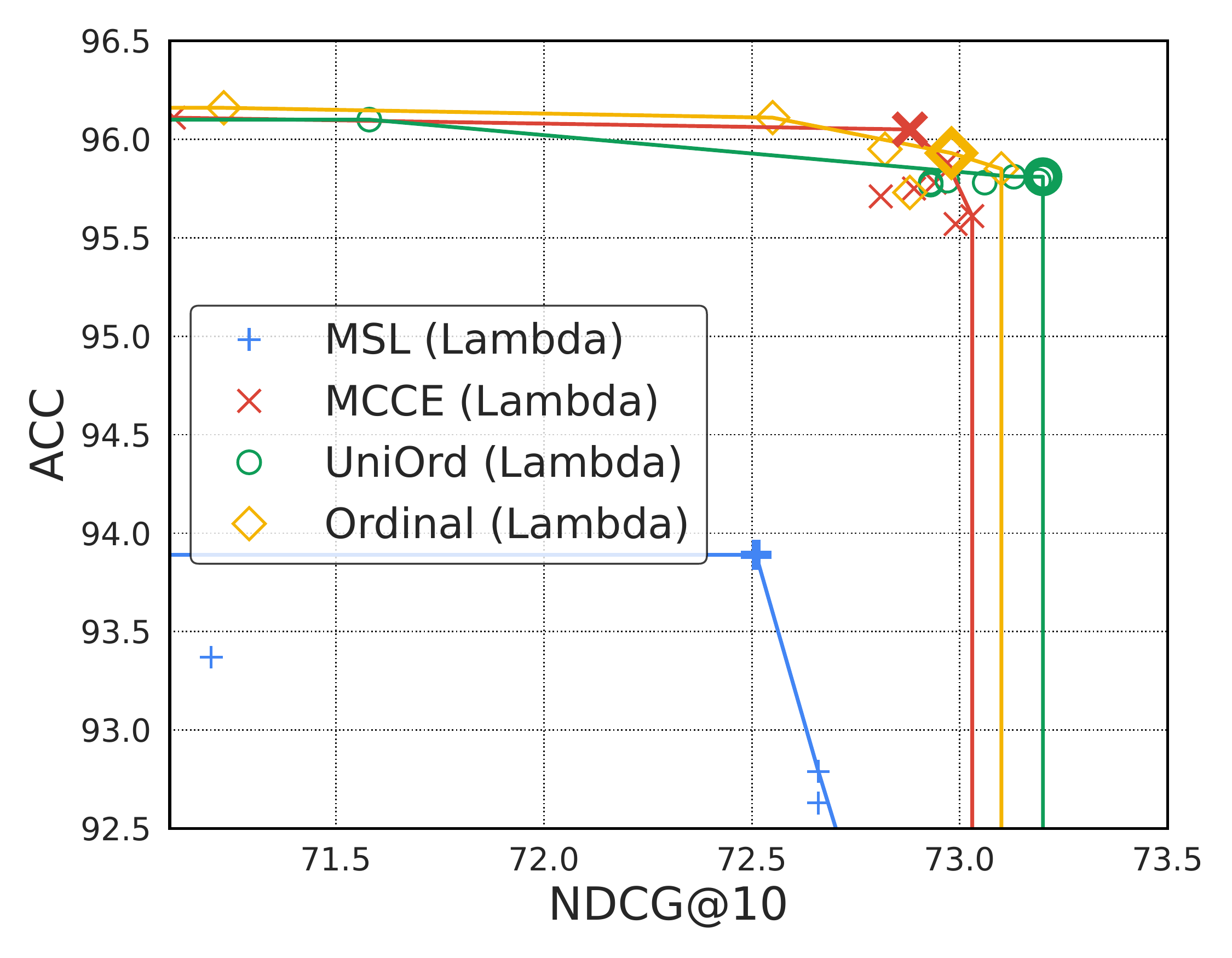}
  \vspace{-0.4cm}
  \caption{Tradeoffs of methods on classification accuracy (ACC) versus NDCG@10. Lines correspond to the Pareto fronts of different grade prediction objectives in the multiobjective setting, labeled in the legend. The results best balancing ACC and NDCG@10, marked in bold, are chosen to represent the MultiObj method in \tblref{tbl:result}.}
  \label{fig:tradeoff}
\vspace{-0.3cm}
\end{figure*}

\paragraph{Metrics}
To quantify the methods on both grade prediction accuracy and ranking, we consider metrics in both categories. 
For ranking performance, we measure NDCG metrics \cite{jarvelin2002cumulated}, which we try to maximize.  Specifically, we use NDCG@10, which scores the top 10 positions. 
For grade prediction performance, we want to minimize cross entropy (CE) in \eqnref{eq:mcce} and the mean square error (MSE) in \eqnref{eq:mse}, and to maximize the classification accuracy (ACC). 
The grade prediction metrics CE and ACC depend on predictions of grade probabilities. 
These are not defined by ranking methods that predict a single score. To evaluate such metrics for ranking methods, we convert ranking scores to grade probabilities by introducing ordinal boundaries $\phi_l$, as those used for univariate ordinal regression \eqnsref{eq:univariate}{eq:uniprob}. The boundaries $\phi_l$ are trained to optimize cross entropy in \eqnref{eq:mcce} with fixed model parameters $\theta$.

\subsection{Results and Discussion} \label{sec:result}

The main results are summarized in ~\tblref{tbl:result}. 
We can make the following observations: 
(\textit{i}) In terms of grade prediction performance, MCCE and Ordinal are strong baselines: they show the best competitive CE and ACC performance, which they are directly optimized for. In addition, by predicting the expected grade value using \eqnsref{eq:mcceexpect}{eq:ordexpect}, they also give competitive MSE.
(\textit{ii}) More interestingly, on Web30K, multiobjective setting combining MCCE objective and Lambda objective shows the best CE, MSE, and ACC. This demonstrates that the ranking objective is synergetic to the grade prediction with MCCE on this dataset.
(\textit{iii}) Similarly, the best ranking NDCG is approached by one of the proposed multiobjective method on each dataset: MCCE (Lambda) on Web30K, Ordinal (Lambda) on Yahoo, and UniOrd (Lambda) on Istella. These best values are statistically significantly better than the state-of-the-art ranking baselines, which also indicates a synergetic interaction of two objectives on the ranking task.
(\textit{iv}) On contrary, the traditional mutliobjective method combining MSL and Lambda show inferior grade prediction performance to MSL only and inferior ranking performance to Lambda (except on Yahoo). This implies no synergy between MSL and ranking losses.

We further analyze the behaviors of the methods in terms of their trade-offs between the ranking performance (measured by NDCG@10) and the grade prediction performance (measured by ACC). The results are shown in \figref{fig:tradeoff}. For each of the multiobjective methods, we can probe multiple points by varying $\alpha$, and we connect the Pareto frontiers for each combination of a grade prediction method and the Lambda method. 
From the tradeoff plot, we observe: 
(\textit{i}) The grade prediction objective and the ranking objective are not simply trading off with each other, but can work collaboratively in certain range of rank weight $\alpha$; 
(\textit{ii}) Proposed combinations of grade prediction objective (MCCE, Ordinal, and UniOrd) and ranking objective probe different Pareto frontiers on different datasets, and are consistently better than a simple combination of MSL and Lambda. These behaviors provide guidance to practitioners: Depending on the dataset, practitioners can bias towards one of multiobjective methods and tune the ranking objective weight $\alpha$ for the best balance of grade prediction and ranking.

As this work focuses on the neural network models, whether these observations could be extended to GBDT models needs further study. But we foresee no constraints to limit the generalization.

\section{Conclusion} \label{sec:conclu}
We provided a rigorous study of learning to rank with graded labels when grades matter, which has practical values but is less explored in the literature. We studied several existing classification and state-of-the-art ranking methods, and proposed several methods by addressing challenges of performing learning to rank with the goal of also accurately predicting ordinal grades. Experiments show that grade prediction and ranking can have synergetic interaction, allowing us to push the Pareto frontier in the ranking and grade prediction trade-off.

\newpage
\bibliographystyle{ACM-Reference-Format}
\balance
\bibliography{references}

\end{document}